# Model for simulating mechanisms responsible of similarities between people connected in networks of social relations

Błażej Żak[1] and Anita Zbieg[2]

[1] Wrocław University of Technology, Wroclaw, Poland
blazej.zak@pwr.wroc.pl
[2] University of Wrocław, Wrocław, Poland
anita.zbieg@gmail.com

**Abstract:** The literature's analysis has identified three social mechanisms explaining the similarity between people connected in the network of social relations: homophily, confounding and social contagion. The article proposes a model for simulating mechanisms responsible for similarity of attitudes in networks of social relations; along with a measure that is able to indicate which of the three mechanisms has taken major role in the process.

## Background

Opinion creation is a social process. Attitudes and behaviors are embedded within a complex system of social interactions and relations with other people. Social psychology in 50's was interested in the subject of social influence in groups, resulting in theories and ideas of conformity (Asch, 1956), social comparison (Festinger, 1954) and research methods e.g. sociometry (Moreno, 1951). In next years social studies focused on the sphere of relations between individuals has not been very popular, especially the one explaining human attitudes and behavior. More popular became the studies of behavior and attitudes where a researcher is focused on characteristics of the individuals. That is why Eagly and Chaiken after extensive literature review pointed out that social context in attitudes' research was insufficiently explored (Eagly, Chaiken, 1993). In the same time the results of studies combining social context and individual's actions in the research were widely discussed, e.g. attitudes versus actions (LaPierre, 1925), Small World Theory and experiment (Miligram, 1967) and concept of The Strength of Weak Ties (Granovetter 1973). The same social context of attitudes and behaviors is recently a research field for Social Network Analysis, focused on relations between people rather than the people themselves. The studies brought new explanations for spread of psychological phenomena such as happiness and solitude, or behavior patterns like smoking or drinking (Christakis N. A., Fowler J. H. 2009). They pointed out that the network of interpersonal relationships among individuals is an important factor that should be observed in social research, enhanced the fact that human behavior or attitudes often spread epidemiologically in the network of connected people. Dynamics of social interactions (Centola D. 2010) and social-economic exchange (Castells M. 2007) are additionally under examination within SNA.

In this article we focus on results and conclusions of SNA analysis based on data from Framingham Heart Study[1] conducted by N. A. Christakis and J. H. Fowler in 2009. Researchers proved that people connected with each other in networks of social relations have similar characteristics. Relations between individuals can be considered as channels for spreading emotions (e.g. happiness), attitudes (e.g. acceptance of sexual behavior in the 60's), and behavior (e.g. eating habits). Moreover, the researchers created the concept of three degrees of social influence. It says that each individual is influenced by their friends ($1^{st}$ degree of separation), but also by friends of friends ($2^{nd}$ degree of separation) and friends of friends' friends ($3^{rd}$ degree of separation).

---

[1] Framingham Heart Study is a research program at Boston University and National Institute of Heart, Blood and Lung, dealing with longitudinal studies, aiming to award the main factors responsible for cardiovascular disease. Extensive questionnaires were handed out since 1948 on large populations. Thousands of people asked in questionnaires were also a source of data for Nicholas A. Christakis and James J. Fowler. http://www.framinghamheartstudy.org [01/07/2010]

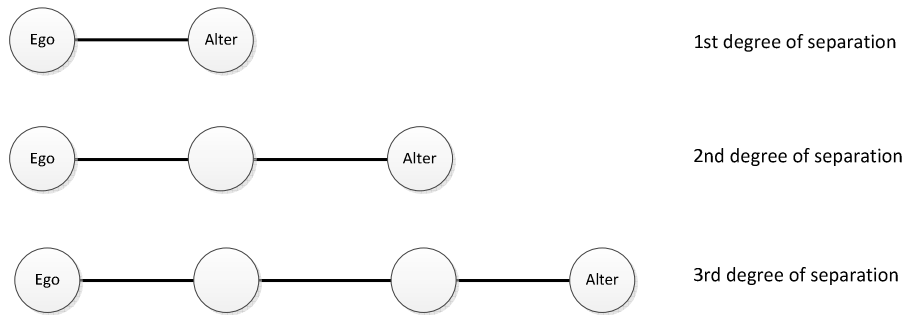

**Diagram 1** – Three degrees of separation

The studies have proven that people connected in network of interpersonal relations tend to demonstrate similar attitudes and behaviors. The more connected people are, the more similar attitudes they should manifest, previously acquired through learning and acting with others (Kotler P. and others, 1999). This phenomenon should manifest itself for concrete and cognitively accessible attitudes (Aronson E., Wilson T. D., Akert R. M. 2009) and between individuals connected in strong relationships. Three mechanisms responsible for similarities between people connected in social network (homophily, social contagion and confounding) have been identified by (Christakis N. A., Fowler J. H. 2009):

## Homophily

> *Homophily is the tendency of individuals to associate with similar individuals, where nodes' similarity affects the formation of relationships between nodes.*

Homophily rule says that similar people attract themselves. Smokers are better understood by other smokers, overweight people feel better in the group of thick friends instead of thin ones. Similarity of consumer attitudes may be the result of the homophily when e.g. they get to know each other on Harley Davidson motorcycle rally, on a pop star concert, or in favorite cafe Starbucks. Consumers connect on a common interest for the brand, and their make relationships upon it. Similarity is also one of the positive factors affecting interpersonal attractiveness (Aronson E., Wilson T. D., Akert R. M., 2009) and may be a good base in further relationship creation.
Homophily connects two nodes sharing a strong attitude with each other. To simulate this phenomenon we randomly choose two nodes out of the network that have a strong attitude and then create a new tie between them.

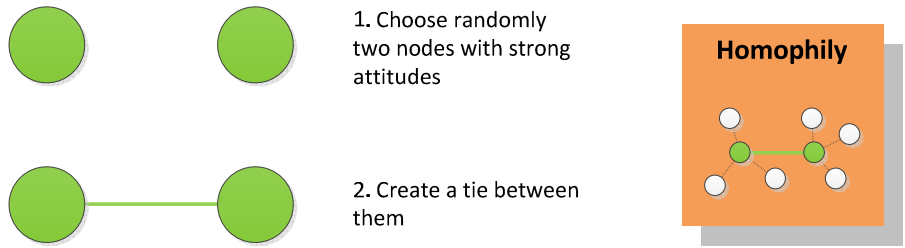

Diagram 2 – Simulation of homophily

## Social contagion

*Social contagion can be interpreted as a mechanism of social impact or influence that spreads across a network of relationships.*

Social contagion can act in two ways: an individual who is the source of influence has a direct effect on another individual (when for example urges to eat greasy sandwiches), or become imitated (e.g. someone begins to eat a lot more for dinner influenced by a colleague eating a lot at the same table). Individual acting as Audi strong consumer gives for example a lift to a cousin and persuades him to buy a car of this brand, or when wanted to have cheap calls to daughter the whole family move to new mobile operator. An interesting conclusion of Christakis and Fowler's research is that the mechanism of social contagion leads to the occurrence of similar attitudes among people connected in the network, however, in contrast to homophily and confounding, this process depends on direction of a relationship between individuals. The study showed that "mutual friends are twice as influential as the friends people name who do not name them back" and "people are not influenced at all by others who name them as friends if they do not name them back" (Christakis – Connected p. 110). Authors give an example: Barry and Kate indicate themselves mutually as good friends, so chance they influence one to another is significant. When Barry indicates Kate as a good friend, but Kate does not indicate Barry, then Kate influences Barry but not vice versa.

Social contagion is a mechanism that spreads across the directed close friends relations. In contrast to information that can spread very quickly throughout a social network, a change in attitudes is a slow process that is a consequence of repetitive interactions between connected people.

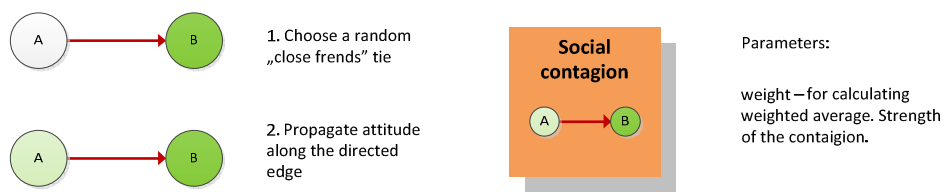

Attitude A = (Attitude B)*weight + (Attitude A)*(1-weight)

Diagram 3 – Simulation of social contagion

The illustration shows social contagious is simulated. Node A connected by outgoing close friends relation get influenced by node B, hence acquiring a little bit of node B attitude. The formula we use for spreading the attitude is a weighted average, with the weight being a parameter of the model. Data collected by Christakis says that the influence is twice as strong for mutual relationships and therefore we double the weight for mutual close friend relations.

## Confounding

*Confounding is a phenomenon describing a simultaneous effect on individuals connected in social networks, where external factor affects people connected in a relationship.*

For overweight friends an external factor influencing them in the past could be a new fast food opened nearby. Friends gained weight, because had regularly visited the new place together. For consumers it can be a TV ad viewed together on a sofa, or a sample of product received by a couple, that has influenced their attitude towards a brand. Thus, confounding creates a similarity in the network, by influencing simultaneously people connected in the network.

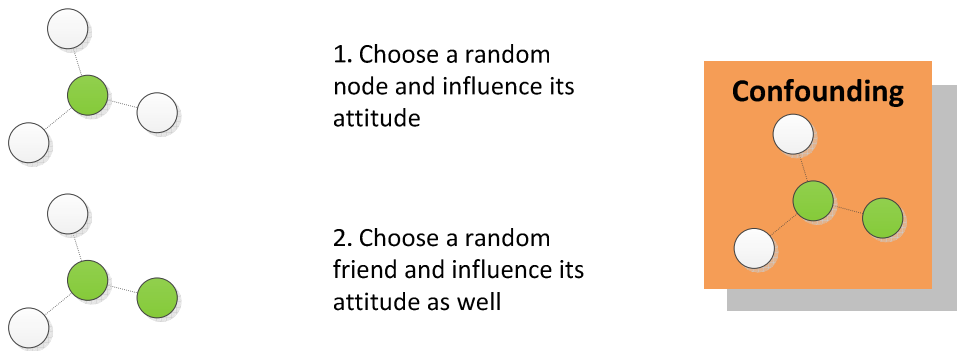

**Diagram 4** – Simulation of confounding

Confounding can be understood as a simultaneous change of attitude of two connected nodes. We can simulate the process in few steps:
1) First select a random node from the network
2) Then select a random friend of the chosen node
3) Influence the attitude of both nodes selected

6## Model & simulation

To further understand the phenomena we propose a generative model for simulation of spread of attitudes across networks of social relations, by simulating the three mechanisms: homophily, confounding and social contagion.

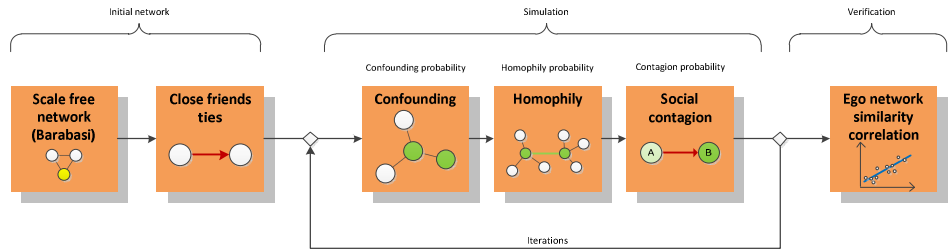

**Diagram 5 –** Model for simulating spread of attitudes

The model overview is presented by the above diagram. It consist of six steps – the first two are responsible for creating a base network with all necessary relations required to make the further simulation. Then the three basic mechanisms responsible for spreading attitudes are simulated within a loop, and afterwards the model is verified by calculating the *ego network similarity correlation*.

## Scale free network

Social networks exhibit scale free characteristics and follow the power law of vertex degree distribution. To simulate spread of attitudes within social network we shall start with a network wired in a proper way - being scale free and following power law.

1. Start with two connected nodes
2. Each new node attach to k existing nodes
3. Probability of making a tie is proportional to number of ties of the existing node (rich get richer)

Parameters:

n – number of nodes to create

K – number of nodes new node gets connected to

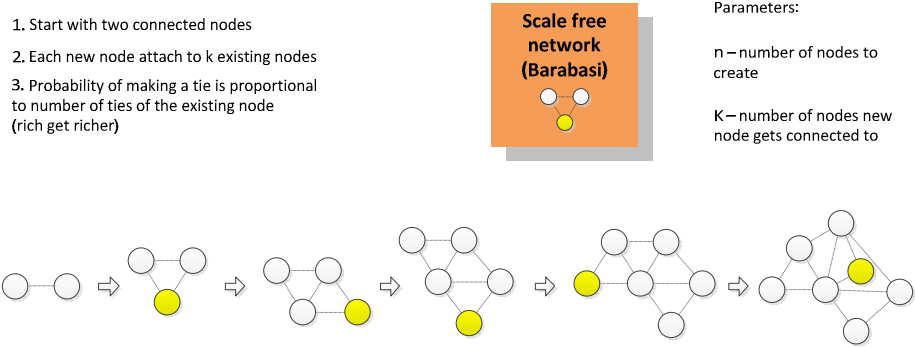

**Diagram 6**- Barabasi's scale free generative model

Barabási proposes a simple generative model for creating an artificial network that have similar characteristics to real life social networks (Barabási i Réka, 1999). The power law and scale free feature of networks is found to be a consequence of the two mechanisms:
1. Continuous addition of new vertices to the network, and
2. Preference of attaching new vertex to already well connected nodes

In the model proposed by Barabási in each step a new vertex is added to an existing network and then attached to *m* existing nodes. The probability of attaching to a specific existing node is proportional to number of ties it already has and might be expressed with a formula:

$$\pi(k_i) = \frac{k_i}{\sum_j k_j}$$

Where $k_i$ is connectivity (degree) of vertex i, and $\pi(k_i)$ is probability of attaching new node to node i.

To be able to simulate social contagion we need to identify which of the connections generated by Barabási model are directional or mutual "close friends" relations. In the real world people were actually asked to point out friends they spend free time with and discuss important issues. In our model we will transform "friends" relations created in a previous step into "close friends" relation. The probability of making such transformation is a parameter of the model as well as proportion between directed and mutual relations to be created[2].

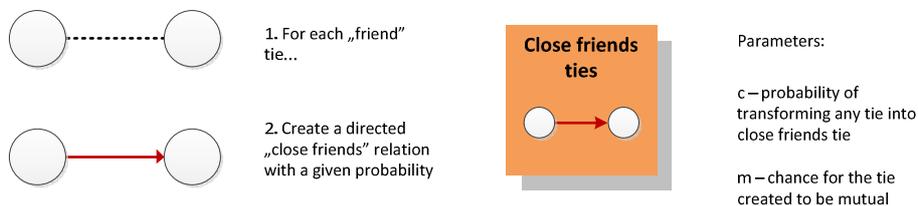

**Diagram 7** – Creation of close friend's ties

---

[2] Research show that an average American usually points out 2-6 influential friends with 5% chance of pointing more than 8 people and 12% chance of not pointing anyone. Christakis NA, Fowler JH, Connected. The Surprising Power of Our Social Networks and How They Shape Our Lives, Little, Brown and Company, New York 2009, p. 48

## Ego-Alter similarity correlation

The last step is verification of the model which is done by calculating the similarity of attitude of a node and an average attitude of node close friends. The verification is done in two steps. First, for each node, we calculate an average attitude of its friends separately for incoming, outgoing and close friend relations. Afterwards a Pearson's correlation is calculated between the node's attitude and the average, separately for each kind of relation. This correlation says how nodes in the network are similar to others they are connected to (first degree of influence). Similarly we can calculate a correlation to check if node is also similar to friends of friends (second degree of influence), and friends of friends of friends.

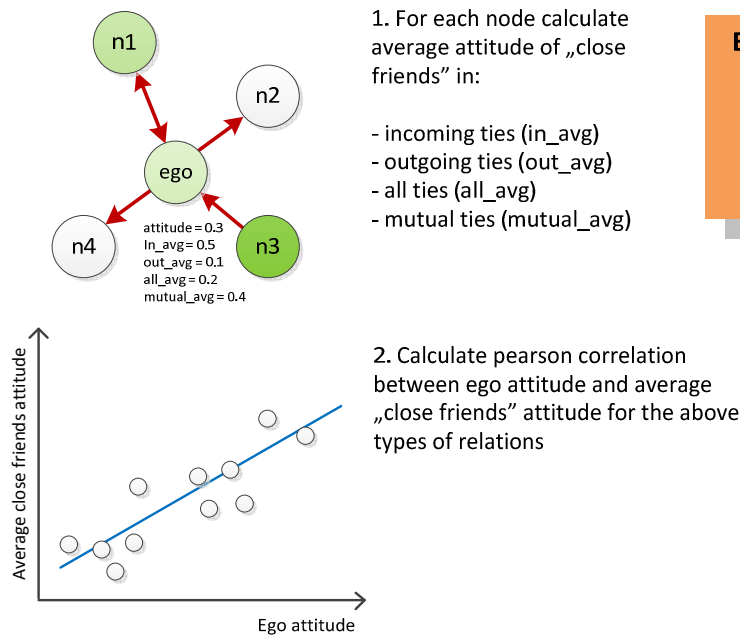
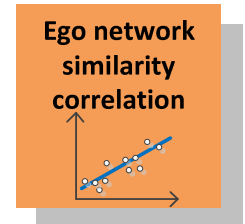

**Diagram 8** – Calculating ego network similarity correlation

To calculate the correlation, first we calculate an average attitude within connected nodes. The average is calculated separately for each kind of relation (incoming, outgoing, mutual), and for each degree of separation (friends, friends of friends, friends of friends' friends). We put results into matrix where columns represent graph nodes, and rows contain average attitudes for each kind of relation. A sample matrix has been presented below.

| Node | W1 | W2 | .. | Wn |
|---|---|---|---|---|
| Ego attitude | 0,07 | 0,19 | .. | 0,31 |
| Average attitude for friends | 0,06 | 0,19 | .. | 0,18 |
| Average attitude for friends of friends | 0,08 | 0,13 | .. | 0,13 |
| Average attitude for friends of friends' friends | 0,10 | 0,11 | .. | 0,11 |
| Average attitude for incoming ties friends | 0,04 | 0,15 | .. | 0,18 |
| Average attitude for outgoing ties friends | 0,08 | 0,19 | .. | 0,15 |
| Average attitude for mutual ties friends | 0,05 | 0,15 | .. | 0,15 |

**Table 1**- Sample data matrix for calculating correlations

Using the matrix we calculate a Pearson's correlation between Ego attitude and Alters for each kind of relation.

## Results

We carried out separate simulations for each investigated mechanism responsible for similarity of people connected with each other in social networks: social contagion, homophily and confounding.

**Social contagion**

Only social contagion is the only mechanism where direction of the relationship plays any role, the other two mechanisms are based on not directional friendship relations. Simulation of this mechanism showed the phenomenon. Charts bellow show correlations for a simulation of contagious attitude spread after 50000 iterations. Indeed in the simulated network mutual friends (Ego ↔ Alter) are more similar (p=0.93) than friends pointed out in one way relationship (Ego → Alter) where correlation is 0.87, and attitudes are the least similar (p=0,79) with friends connected with Ego by an incoming relationship (Ego ← Alter).
Charts bellow show correlations for a simulation of contagious attitude spread after 50000 iterations. In the generated network mutual friends are more similar than friends pointed out in one way relationship, and attitudes are the least similar with friends connected by an incoming relationship.

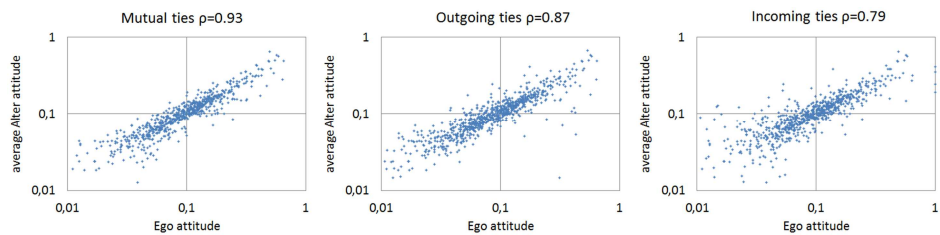

**Diagram 9** – Ego and Alter attitudes for different types of ties (incoming, outgoing, mutual) after 50000 steps of social contagion mechanism simulation

It is possible also to observe the three degrees of influence rule proposed by Fowler and Christakis. The ego network similarity correlation is observable up to third degree of friends. Whereas nodes are very similar to their friends (strong correlation), they are still but less similar to friends of their friends (medium correlation). In the third degree of separation (friends of friends of friends) the similarity is still observable but very low (correlation 0.16), and it disappears completely when we move one more step from the node.

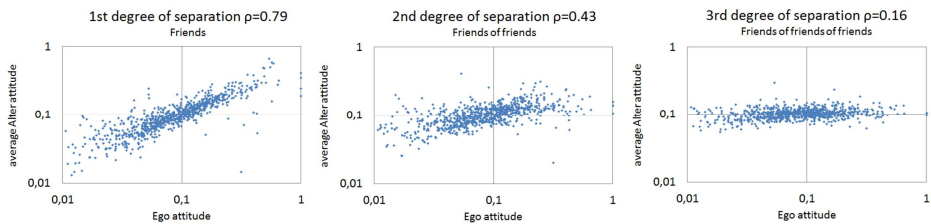

**Diagram 10 – Ego and Alter attitudes for different degrees of separation**

The model proposed is stochastic, which means it is largely dependent upon a random variable. Nevertheless results obtained in independent simulations are repetitive and demonstrate the same characteristic, what is presented on the following diagrams.

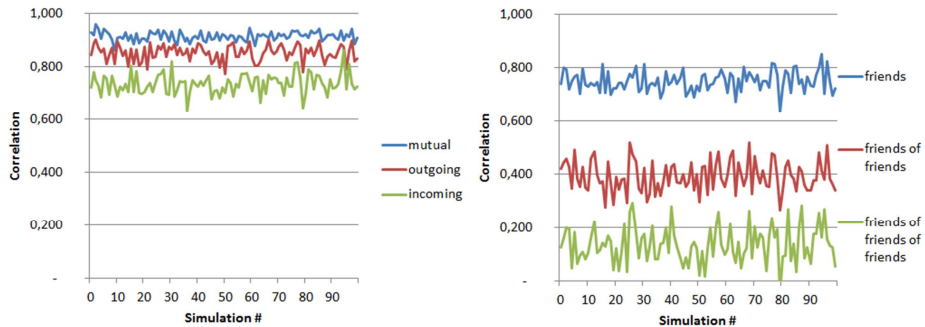

**Diagram 11** – Ego network correlations for 100 repetitive independent simulations of social contagion mechanism

With the model we can observe how attitudes within the Ego network are becoming similar after each iteration of social contagion simulation process. The graph bellow presents correlations between Ego and Alter attitudes for different type of ties. The X axis represents algorithm iteration, in other words we can observe how correlations change in time.

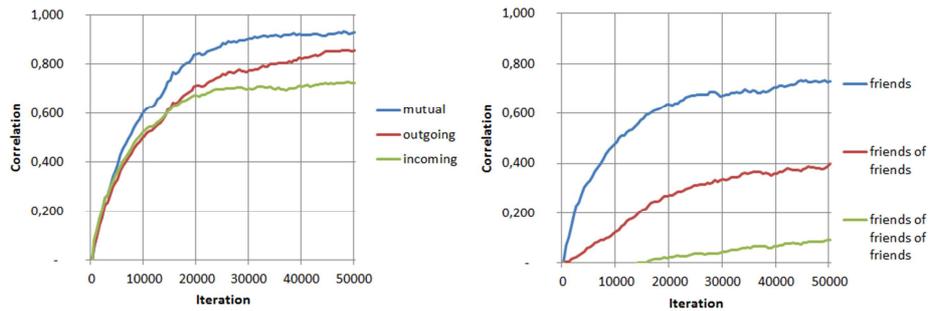

**Diagram 12** - Ego network correlations for contagious attitude in the successive iterations.

We can notice that at the beginning the graph illustrates fully random distribution of attitudes, but in further steps of simulation nodes become more and more similar: Ego attitude becomes more similar to attitudes of friends, but also to attitudes friends of friends' and Alters in 3rd degree of separation. We can also observe the strongest similarity for friends mutually connected, weaker correlation for Alters that Ego indicated as a friends. The weakest pattern is between Ego attitude and attitudes of Alters indicating Ego as a friend, but still significant.

**Homophily**

When we simulate homophily as a dominant mechanism, the correlations take similar values for each type od of relationships: incoming, outgoing and mutual. The graph of similarity for each degree of separation is similar to the one presented in social contagion with one exception explained below. The following graphs show the correlation calculated for one hundred independent simulations.

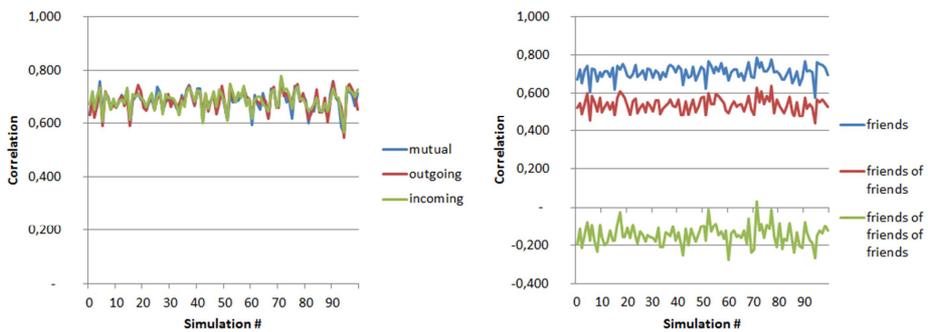

**Diagram 13 -** Ego network correlations for 100 repetitive independent simulations of homophily mechanism

It's interesting that for homophily correlation between Ego attitude and attitude presented by friends of friend of friends is negative. In further investigation presented on the next graphs this phenomenon is visible better.

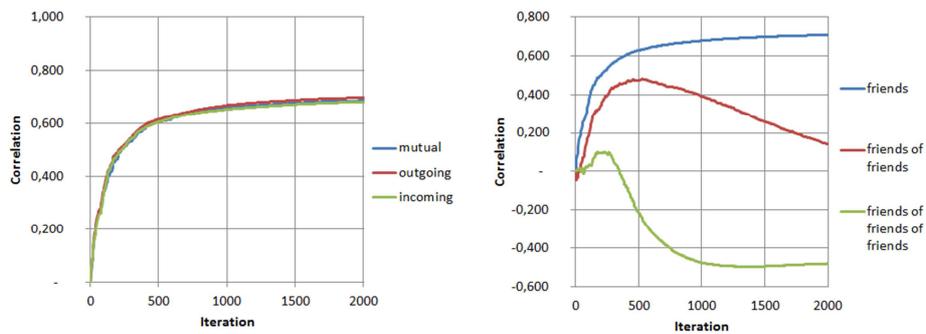

**Diagram 14** - Ego network correlations in the successive iterations of homophily mechanism.

We can observe that in first steps of simulation, similarity between Ego and Alters increases for $1^{st}$, $2^{nd}$ and $3^{rd}$ degree of separation. Then the similarity to Ego starts decrease slightly for friends of friends and collapse rapidly for friends of friends of friends to p=-0,5. It means that Ego is rather different, than similar to people from $3^{rd}$ degree of separation. To understand this phenomenon, we visualize the network created along the simulation.

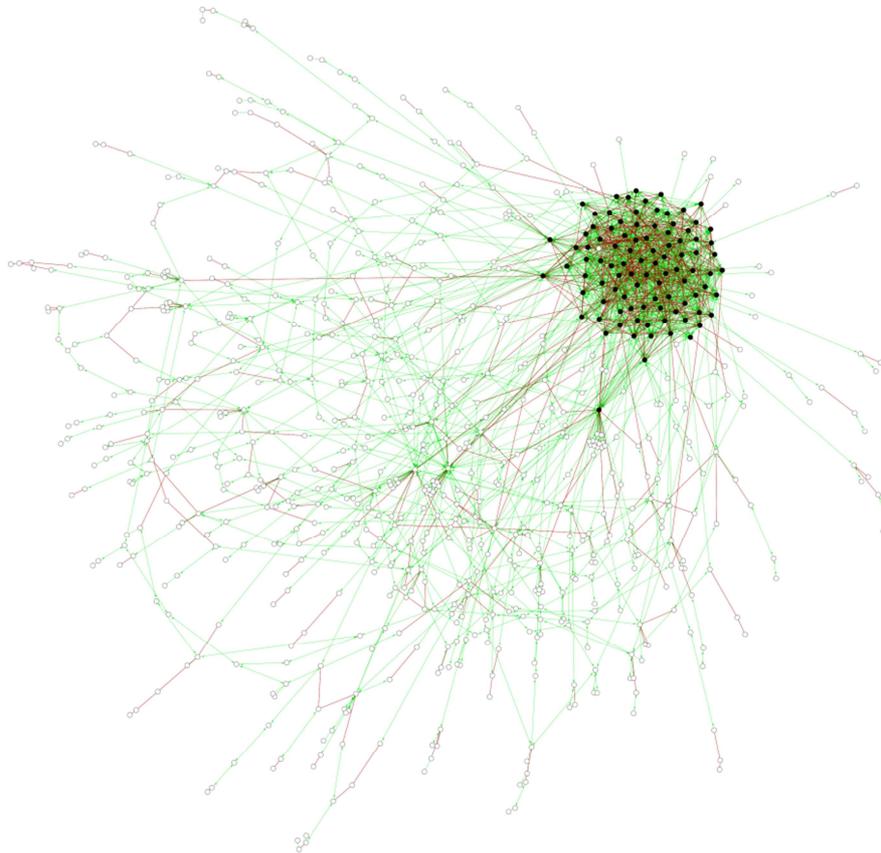

**Diagram 15** – Visualization of a graph created by simulating homophily mechanism

In the structure of the graph we clearly see a clique formed by people that share the same attitude. With long-acting homophily mechanism the clique connects people so densely that in most cases for Ego belonging to a clique, Alters from 3rd degree of separation are beyond the clique. Hence, the correlation between Ego and friends of friends of friends is negative – Ego is different from these Alters, because they don't belong to the clique.

### Confounding

The correlation coefficient is the least sensitive for the confounding mechanism, even though it is also simulated as a networked mechanism. Each time affecting simultaneously on two connected persons, we might expect to observe emerging similarity between Ego and Alter. The simulation does not confirm this hypothesis. It shows that confounding mechanism simulated in our model is not able to make network nodes similar. In fact, the correlation calculated over one hundred independent simulations were close to 0 and never outnumbers 0.2 (see Diagram 16 below).

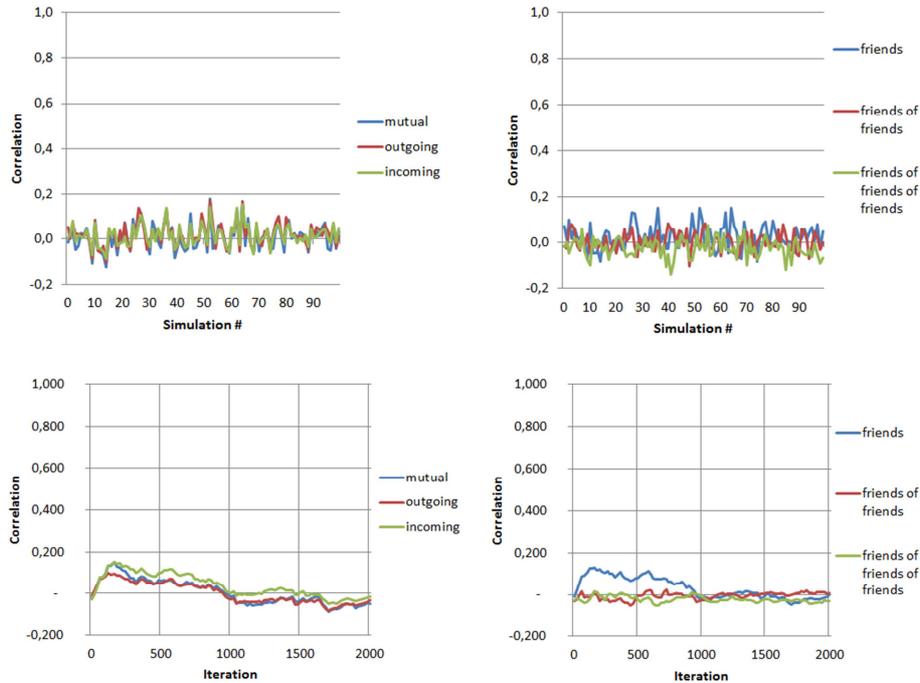

**Diagram 16** - Ego network correlations for 100 repetitive independent simulations of confounding mechanism in the successive iterations

## Conclusions

The differences in correlations are observable across multiple networks generated with the model, even though their actual values might vary due to randomness of the simulation process. Furthermore the correlations reflect which mechanism (homophily, confounding or social contagion) has played a major role in the simulation. Contrary to the above example if we simulate strong homophily without the contagion the nodes still get similar up to the third degree, but correlations do not differ for each type of relationship (incoming, outgoing, mutual). Confounding as a mechanism on the other hand generate networks with much smaller correlations than homophily and contagion. Ability to determine which of the mechanisms have taken major role in developing similarities between individuals connected in the network is critical for understanding how attitudes and behaviors spread across networks of social relationships. Anyway further analysis and empirical studies should be made to fully prove validity of the model.